\input pah0.eps

\documentstyle[11pt,GAT65]{article}

\begin{document}

\title{Models for Type Ia Supernovae and Cosmology}
\author{Peter H\"oflich}
\affil{Department of Astronomy, University of Texas, Austin, TX 78681, USA}

\begin{abstract}
 From the spectra and light curves it is clear that SNIa events
are thermonuclear explosions of white dwarfs. However, details
of the explosion are  highly under debate. Here, we present detailed models
which are consistent with respect to the explosion mechanism,
the optical and infrared light curves (LC), and the spectral evolution. This
leaves the
description of the burning front and the structure of the white dwarf as
the only free parameters. The explosions are calculated using  one-dimensional
Lagrangian codes including nuclear networks. Subsequently, optical and IR-LCs
are
constructed. Detailed NLTE-spectra are computed for several instants of time
using
the density, chemical and luminosity structure resulting from the LCs (sect.2).

 Different models for the thermonuclear explosion are discussed including
detonations, deflagrations, delayed detonations, pulsating delayed detonations
(PDD)
 and helium detonations (sect.3).
Comparisons between theoretical and observed LCs and spectra provide
an insight into details of the explosion and nature of the progenitor stars
(sect. 4 \& 5).
 We try to answer several related questions.
 Are subluminous SNe~Ia a group different from `normal' SN~Ia (sect. 4)?
  Can we understand observed properties of the LCs and spectra
(sect. 5)? What can we learn from infrared spectra?
Can we determine $H_o$ independently from primary distance indicators, and
how do the results compare with empirical methods (sect.6)?
 What do we learn about the progenitor evolution and its metallicity?
 What are the systematic effects for the determination of
the cosmological parameters $\Omega_M$
and $\Lambda$ and how can we recognize this potential 'pitfalls' and correct
for
evolutionary effects (sect.7)?

\end{abstract}


\section{Introduction}
During the last few years it became evident that Type Ia supernovae
are a less homogeneous class than previously believed (e.g. Barbon et al.,
1990,
Pskovskii 1970, Phillips et al. 1987).
 In particular,
the observation of subluminous SN1991bg
 (Fillipenko et al. 1992, Leibundgut et al. 1993)
 raised questions on the SN-rate   and whether we have missed a huge subgroup.
The possible impact on our understanding of supernovae statistics and,
consequently, the chemical evolution of galaxies must be noted.

With respect to the use of Type Ia Supernovae as distance indicators,
for nearby supernovae ($z \leq 0.1$),
 different schemes have been developed and tested
to cope with the problems of deducing the intrinsic brightness
 based on theoretical models or observed correlations between spectra or light
curves and the
absolute brightness using primary distance indicators (e.g. Norgaard-Nielsen et
al. 1989;
 Branch \& Tammann 1992,  Sandage \& Tammann 1993,  M\"uller \& H\"oflich 1994,
Hamuy et al. 1996; Ries, Kirshner \& Press 1995; H\"oflich \& Khokhlov
1996, HK97 thereafter). These methods have been tested locally and provide
consistent results.
 Due to the new recalibrations of the brightness of SNe~Ia by $\delta $Cephii
in nearby galaxies
(e.g. Saha et al. 1997 and references therein),
a common agreement could be reached on the value of $H_o$.
In a next step, new telescopes and observational
 programs put SNe~Ia at large red shifts well within reach and justify
optimism for the discovery of a large number of distant SNe~Ia.
The Berkeley group has discovered more than 50 SNe~Ia
up to a red shift of 0.9 (Pennypacker et al. 1991, Perlmutter et al. 1997;
Pennypacker, private communication). The
 CfA/CTIO/ESO/MSSSO collaboration  has found a similar number of
supernovae (Leibundgut et al. 1995, Schmidt et al. 1996). A major
concern with the use of SNe~Ia to determine $q_o$ is  systematic
errors due to evolutionary effects (H\"oflich et al. 1997). This is especially
true for statistical methods  which are calibrated only on local SNe~Ia.

Despite the success of purely empirical methods, theoretical work  on SNe~Ia is
critical
to provide an independent test for the distance scale, to get an insight into
the
underlying physics of the explosion, it provides a tool to investigate benefits
of new wavelength
ranges not yet well explored (i.e. the IR) and it allows to test for the
importance
(or unimportance) of systematic effects when it comes to the study of very
distant SNe~Ia.
The goal of this paper is to provide a brief overview of the different aspects
just mentioned.

\section{Numerical Methods}
 A consistent treatment of the explosion, light curves and spectra is critical
to provide a
tight coupling between the explosion model and the observational quantities.
Therefore,
we start with a brief description of the numerical methods.

\subsection { Hydrodynamics}

The explosions are calculated using a one-dimensional radiation-hydro
code, including nuclear networks (HK96 and
references therein).  This code solves the hydrodynamical equations
explicitly by the piecewise parabolic method (Collela \& Woodward 1984)
and includes the solution of the frequency averaged radiation transport
implicitly via moment equations, expansion opacities, and a detailed
equation of state. The frequency averaged variable Eddington factors
and  mean opacities are calculated by solving the frequency dependent
transport equations.
About one thousand frequencies (in one hundred frequency groups) and
about five hundred depth points are used.  Nuclear burning is
taken into account using a network which has been tested in many
explosive environments (see Thielemann, Nomoto \& Hashimoto 1996 and
references therein).

\subsection { Light Curves}

Based on the explosion models, the subsequent expansion
 and  bolometric as well as monochromatic LCs
 are calculated using a scheme recently developed, tested
and widely applied to  SN Ia (e.g. H\"oflich et al. 1993, H\"oflich et al.
1997ab and references therein).
The code used in this phase is similar to that described above, but
 nuclear burning is neglected and
 $\gamma $ ray transport is included via a Monte Carlo scheme (H\"oflich,
M\"uller \& Khokhlov 1992).
In order to allow for a more consistent treatment of scattering, we
solve both the (two lowest) time-dependent radiation moment equations for the
radiation
energy and the radiation flux, and a total energy equation.
At each time step, we then use $T(r)$ to determine the
Eddington factors and mean free paths of photons by solving the
frequency-dependent
radiation transport equation in comoving frame (see next section) in about 100
frequency bands (see below)
and integrate to obtain the frequency-averaged quantities.  We  use the
latter to iterate the solution with the frequency-integrated
energy and flux equations.
 The frequency averaged  opacities, have
been calculated under the assumption
of local thermodynamical equilibrium. This is a reasonable approximation
for the light curve  since diffusion time scales are always governed by layers
of large optical depths.
 Note that the comparison of L(r) between the frequency independent
solution and the  frequency dependent solution provides a critical test for the
consistency of
the approximations used in the radiation hydro.

Both the monochromatic and mean opacities are calculated using the Sobolev
approximation
(Sobolev 1957) to
calculate the absorption probability within a shell and to include line
blanketing. The approach
is similar to Karp et al. (1977) but generalized for the comoving frame and the
integration boundaries are
adjusted to a radial grid (H\"oflich 1990).
The scattering, photon redistribution  and thermalization terms
used in the light curve opacity calculation are calibrated with NLTE
calculations
using the formalism of the equivalent-two-level approach
(H\"oflich 1995).

To calculate the monochromatic light curves, we use the $T(r)$
with  the time dependence of the structure given by the frequency-integrated
solution of the momentum equations
 to solve  the frequency-dependent transfer equations in LTE every 0.2 to 0.5
days
solution  to get $L_\nu$ in the observer's frame.
 The broad band light curves are determine by convolution
of $L_\nu$ with the filter functions.
We use a few hundred
frequency bands with the same scheme as described in the following subsection.
 Note that a proper treatment of the frequency derivatives in the comoving
frame
equations requires the use of about 5 to 10 times more frequencies than
frequency bands.
 To test the consistency
between the frequency-dependent and frequency-independent calculations
we also integrate  the $L_\nu$  and check to see how close the resulting L
is to that from the solution of the frequency-integrated moment equations. The
solutions
are consistent within 10 \%  (see e.g. Khokhlov et al. 1993, H\"oflich et al.
1997b).
 Monochromatic colors from our light curve code have been compared to colors
calculated
by our detailed NLTE spectral code (H\"oflich 1995).
 Based on this comparison, the solutions for the Type Ia models are good
to better than 0.1 mag near maximum light and deteriorate to 0.3 mag at
about 100 days. Note that these tests probe the internal consistency only.
However,
a direct comparison of the bolometric correction of observered supernovae and
of the
theoretical models provide an independent test for systematic errors and, in
particular,
for the (critical) translation of bolometric into monochromatic brightness.
 For the well observed supernovae SN~1992a, accurate BCs have been
reconstructed and found
to be consistent with our models  within $0.1^m$ (HK96).
 According to the certainties, uncertainties in the rise times are about 1-2
days due
to the flat maxima.

\subsection {NLTE-Spectra}

Finally, detailed NLTE spectra have been constructed based on
the  LC calculations.
Thus, the effect of energy stored during previous epochs is properly taken into
account.
 The energetics of the model are calculated. Given an explosion model, the
evolution of
 the spectrum is not subject to any tuning or free parameters such as the total
luminosity.
 A modified version of our  code for Nlte Extended ATmospheres (NEAT) is used.
Although time dependence in the rate equations can be included in this code, it
was found to be
 For details  see   H\"oflich (1990, 1995), H\"oflich et al. (1997b) and
references therein.

For the NLTE spectra,
 the  density, chemical profiles and the luminosity
as a function of the radial distance $r$ are given by  the hydrodynamical
explosion and  the
 LC calculations, including  the Monte-Carlo scheme for  $\gamma $-ray
  transport.

\begin{figure}
\psfig{figure=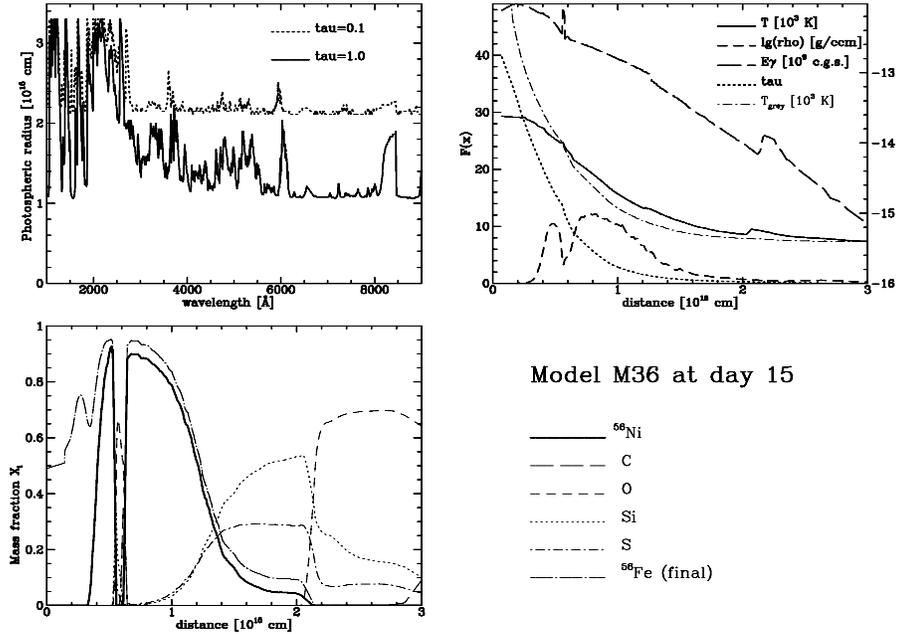,width=12.6cm,rwidth=9.5cm,angle=270}
\caption{ Various quantities for M36 15 days after the explosion.
     Distances  are given as a function of wavelength at which
the monochromatic optical depth reaches 0.1 and 1, respectively
 (upper left),
          Temperature T, density $\rho $, energy deposition due to radioactive
decay $E_{\gamma} $ and Rosseland optical depth are given as a function of
distance
 and, for comparison, $T_{grey}$ for the grey extended atmosphere (upper
right),
and chemical profile are presented for the most abundant elements
(lower).}\label{pah-1}
 \end{figure}

  The radiation transport equation is solved in the comoving frame
including relativistic terms.
For the determination of the first moment of the intensity, we solve
the radiation transport equation for strong lines including `quasi' continua
 (weak lines are treated in a Sololev approximation, see below)
 in the comoving frame, i.e.
$$
 Op~I = \chi (S  -  I)
$$
$$
I = I (\mu, r, \nu);
{}~~~~~~~~~S = S ( r, \nu);~~~~~~~~
\chi = \chi ( r, \nu)
$$
($\mu $: cosine between the radial direction and I; r : radial distance;
v(r) : radial velocity; S: source function; $\chi $: absorption
coefficient; $\nu $: frequency).
       The operator $Op$ can be written as follows (Castor, 1974):

$$
%\eqalign
Op = \Biggl\lbrack\mu +{v(r) \over c}\Biggr\rbrack
 {\partial \over \partial r}\\
+\Biggl\lbrack {(1 - \mu^2) \over r}
 \Bigl(1+ {\mu v(r) \over c} (1-\beta(r))\Bigr)
\Biggr\rbrack      {\partial \over \partial \mu} $$
$$-\Biggl\lbrack {v(r) \nu \over c \cdot r} \Bigl(1 -
\mu^2 + \mu^2 \beta(r)\Bigr) \Biggr\rbrack
{\partial \over \partial \nu} \\
+ \Biggl\lbrack  {3 v(r) \over c \cdot r}
\biggl( 1-\mu^2+ \beta(r) \mu^2 \biggr)
 \Biggr\rbrack\\
$$

with
$$
\beta : = {d~ln~v(r) \over d~ln~ r}.
$$
 The term in front of $  {\partial \over \partial \nu} $ can be
interpreted as the classical Doppler shift. The second terms of the
partial derivatives with respect
to r, $\mu $ and the logarithmic derivative of v(r)
  correspond to the advection and aberration effects.
 For practical purpose, we solve the non relativistic transport equation using
a
Rybicki scheme (Mihalas et al., 1975, 1976ab).
 Overlapping lines in the comoving frame
 cannot be treated because the system is
solved as a boundary problem in the frequency space which, from the concept,
disallows
a propagation of information toward higher/lower frequencies for
expanding/collapsing
envelopes. The same limitation should also apply to Nugent et al. (1995a) who
use the same
method.
However, the problem of overlapping lines can be and has been solved (H\"oflich
1990)
 in full analogy with
the problem of partial redistribution using a Fourtrier-like scheme (Mihalas et
al, 1976b).
Although our code can deal with partial redistribution, this effect is
neglected to save CPU-time.
 A technical problem is the fact that
each line must be sampled by  at least 10  frequency points to provide a
sufficient accurate
representation of the $\delta /\delta \nu$-term (Mihalas et al. 1975).
Currently, we must limit
this approach to  $\approx 10,000 $ lines.

 Blocking by lines other than the strong ones  is included in
   a `quasi' continuum approximation, i.e. the frequency derivative terms in
the radiation
transport equation is included in the narrow line limit  to calculate
the probability for photons  to pass a radial sub-shell along a given direction
$\mu $
(Castor, 1974,  Abbot and Lucy  1985, H\"oflich, 1990).
 Note that the information on the exact location of the interaction within a
subshell gets lost
translating  into an uncertainty in the wavelength location of $\approx
10^{-3}$.

   The statistical equations are solved consistently  with the radiation
transport
 equation to determine the non-LTE occupation numbers using both an accelerated
lambda
iteration (see  Olson  et al.  1986)
         and an  equivalent-two level approach for
 transitions from the ground state which provides an efficient way to take the
non-thermal
fraction of the source function into account during the radiation transport
and, effectively,
accelerate both the convergence rate and the stability of the system. A
comparison of the
    explicit with the implicit source functions provides a sensitive tool to
test for
convergence of the system of rate and radiation transport equations.

 Excitation by     gamma rays
is included. Detailed atomic models are used for up to three most abundant
ionization stages
of several elements, i.e. (He), C, O, Ne, Na, Mg, Si, S, Ca, Fe taking into
account
20 to 30 levels in the main ionization stage. Currently, we use
detailed term-schemes for C, O, Ne, Na, Mg, Si, S, Ca, Fe, Co and Ni.
 The corresponding lower and upper ions are represented by the ground states.
 The energy levels and cross sections of  bound-bound transitions are taken
from
 Kurucz (1993).
A total of 10,000 lines are treated in full NLTE.  In addition, $\approx
1,000,000$ lines out of
a list of 31,000,000 (Kurucz, 1993) are included for the radiation transport.
 For these lines, we assume
LTE population numbers inside each ion. To calculate the ionization balance,
 excitation by   hard radiation is taken into
 account.   LTE-line scattering is included using  an equivalent-two-level
approach calibrated by our NLTE-elements.

The need for  spectral analyses consistent with respect
 to the explosion mechanism,
$\gamma-$ray transport and LC calculation is obvious from Fig.~\ref{pah-1}
No well defined photosphere exists.
 Already at maximum light, the chemical distribution of elements does not
follow
the density profile but shows large individual variations in the
line forming region, i.e. between 1 and 2 $10^{15}~cm$.
The energy deposition
and excitation due to $\gamma $-rays becomes important within the photosphere.
Consequently, the luminosity cannot be assumed to be depth independent
for the  construction of synthetic spectra.

\section{Hydrodynamical Models}

\subsection{Explosions of Massive White Dwarfs}

 A   first group consists of massive carbon-oxygen white
dwarfs (WDs) with a mass close to the Chandrasekhar mass which accrete mass
through Roche-lobe overflow from an evolved companion star (Nomoto \&
Sugimoto 1977).  The explosion is
triggered by compressional heating.
                                 The key question is
how the flame propagates through the white
dwarf. Several models of SNeIa have been proposed in the
past, including detonation (Arnett 1969; Hansen \& Wheeler 1969),
deflagration (e.g. Nomoto, Sugimoto \&
Neo 1976) and the delayed detonation model,
which assumes that the flame starts as a deflagration and turns into a
detonation later on (e.g. Khokhlov 1991ab, Woosley \& Weaver 1994).
\begin{table}
\caption{
Some quantities are given for our models   (see text).
}\label{pahtbl-1}
\begin{center}\scriptsize
\begin{tabular}{llllllll}
\tableline
  Model~~~~~ &  $M_\star$ ~~~~~& $\rho_c$ ~~~~~~~~~ & $\alpha$ ~~~~~&
  $\rho_{tr}$~~~~~~~~~~~~
  & $M_{Ni}$~~~~~  \\
& $[ M_\odot ]$ & $[10^9 c.g.s]$~ &  &   $[ 10^7 c.g.s]$~ &  $[ M_\odot ]$ &
 \\
\tableline
  DET1   &   1.4  &  3.5   &   ---  &  ---  &
                          0.92  \\
DF1    &   1.4  &  3.5   &  0.30  &  ---  &
                          0.50  \\
W7     &  1.4  &  2.0   &  n.a.  &  ---  &
                          0.59  \\
M35     &  1.4  &  2.8   &  0.03  &  3.0  &
                           0.67  \\
M36     &   1.4  &  2.8   &  0.03  &  2.4  &
                           0.60  \\
M37     &   1.4  &  2.8   &  0.03  &  2.0  &
                           0.51  \\
M39     &   1.4  &  2.8   &  0.03  &  1.4  &
                           0.34  \\
DD200c & 1.4  &  2.0   &  0.03  &  2.0  &
                           0.61  \\
DD13c (1:1)   &  2.6   &  0.03  &  3.0  &  1.36  &
                           0.79  \\
DD21c (1:1)   &  1.4  &  2.6   &  0.03  &  2.7  &
                           0.69  \\
DD23c (2:3)   &  1.4  &  2.6   &  0.03  &  2.7  &
                           0.59  \\
PDD3    &    1.4  &  2.1   &  0.04  &  2.0  &
                           0.49  \\
PDD5    &                 1.4  &  2.7   &  0.03  &  0.76 &
                           0.12  \\
PDD8    &                 1.4  &  2.7   &  0.03  &  0.85 &
                           0.18  \\
PDD7    &                 1.4  &  2.7   &  0.03  &  1.1  &
                           0.36  \\
PDD9    &                 1.4  &  2.7   &  0.03  &  1.7  &
                           0.66  \\
PDD6    &                 1.4  &  2.7   &  0.03  &  2.2  &
                           0.56  \\
PDD1a   &                 1.4  &  2.4   &  0.03  &  2.3  &
                           0.61  \\
PDD1c   &                 1.4  &  2.4   &  0.03  &  0.71 &
                           0.10  \\
HeD2    &                 0.6+0.22  &  .013   &  ---   &  ---  &
                           0.43  &   \\
HeD4    &                 1.0+0.18  &  .150   &  ---   &  ---  &
                           1.07  &   \\
HeD6    &                 0.6+0.172 &  .0091  &  ---   &  ---  &
                           0.252 &   \\
HeD10   &                 0.8+0.22  &  .036  &  ---   &  ---  &
                           0.75  \\
HeD12   &                 0.9+0.22  &  .083  &  ---   &  ---  &
                           0.92  \\
 DET2     &               1.2        &  0.04  &  ---  &  ---  &
                                       0.63  \\
DET2env2 &               1.2 + 0.2  &  0.04  &  ---  &  ---  &
                                      0.63  \\
DET2env4 &               1.2 + 0.4  &  0.04  &  ---  &  ---  &
                                      0.63  \\
\tableline
\end{tabular}
\end{center}
\end{table}

\begin{figure}
\psfig{figure=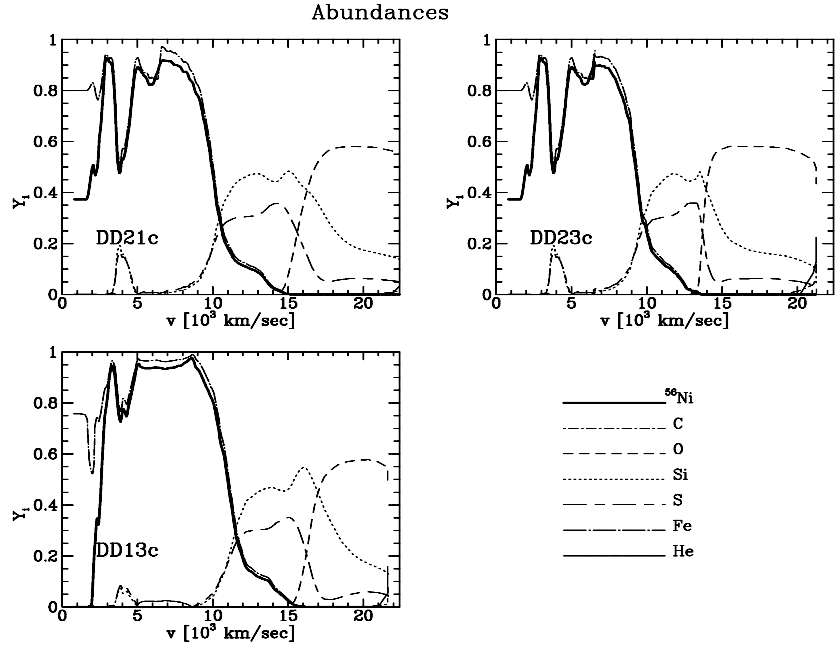,width=12.6cm,rwidth=8.7cm,angle=270}
\caption{
Abundances as a function of the expansion velocity for
three delayed detonation models (see Table~\ref{pahtbl-1}).}\label{pah-2}
 \end{figure}
Our sample includes
     detonations (DET1/2), deflagrations (W7, DF1),
 delayed detonations (M35-39, DD13-27, DD200) and
 pulsating delayed detonations (PDD1-9).
 The deflagration speed is parameterized as $D_{def} = \alpha a_s$, where $a_s$
is the local sound velocity
ahead of the flame and $\alpha $ is a free parameter. The speed of the
detonation wave is given by the
sound-speed behind the front. For delayed detonation models, the transition to
a detonation is given
by another free parameter $\rho _{tr}$. When the density ahead of the
deflagration front
reaches $\rho_{tr}$, the transition  to a detonation is forced by increasing
$\alpha $ to 0.5 over
5 time steps bringing the speed well above the Chapman-Jouguet theshold
for steady deflagration.
 For pulsating delayed detonation models,
 the initial phase of burning fails to release sufficient energy to disrupt the
WD.
During the subsequent contraction phase, compression of the mixed layer of
products of burning and C/O formed
at the dead deflagration front would give rise to a detonation via
compression and  spontaneous ignition (Khokhlov 1991ab).
 In this
scenario, $\rho_{tr} $ represents  the density at which
the detonation is initiated after the burning front dies out.
 Besides the description of the burning front,
 the central density of the WD at the time of the explosion is another free
parameter. For white
dwarfs close to the Chandrasekhar limit, it  depends sensitively on the
chemistry and the accretion rate $\dot M $ at the time of the explosion.

In addition, the C/O ratio has been varied. In general,
it has been assumed to be 1:1 unless otherwise quoted in brackets
after the name. A lower C/O ratio can be expected if the WD originates from a
progenitor
of more than $\approx 3 M_\odot$.
 The initial metallicity for $Z\geq 20$ is assumed to be solar,
but for DD24c, DD25c, DD26c and DD27c for which 1/3, 3, 0.1 and 10
times solar abundances are used, respectively. Otherwise, these models are
identical to DD21c.

 As example, typical chemical structures are given in Fig.~\ref{pah-2}.
Since the burning time
scales to NSE or partial NSE are shorter than the hydrodynamical time
scales for all but the very outer layers, the final products depend
mainly on the density at which burning takes place.
With decreasing
transition density, lesser $^{56}Ni$ is produced and the intermediate
mass elements expand at lower velocities
 because the later
transition to a detonation allows for a longer
pre-expansion of the  outer layers (DD21c vs. DD13c). Similarily,
with increasing C/O ratio in the progenitor, the specific energy
release during the  nuclear burning is reduced (DD23c vs. DD21c) and
the transition density at the burning front reached later in time,
resulting in a larger preexpansion of the outer layers. This
may  allow to determine the main sequence mass of the progenitor.

 To test the influence of the metallicity (i.e. nucleii beyond Ca) we
have constructed models
with parameters identical to DD21c but initial metallicities
between 0.1 and 10 times solar. The energy release,
the density and velocity  structure are virtually identical to that of DD21c.
The main difference is an increase  of more neutron-rich Fe group nuclei,
namely $^{54}Fe$, in the outer layers because, there, $Y_e$ is inherented from
the progenitor.
While, for
1/3 solar metallicity, hardly any $^{54}$ Fe  is produced at high velocities,
$^{54}$Fe is  as high as 5 \% by mass fraction if we start with ten times the
solar metallicity (see sect. 7).
 Note that the temperature in the inner layers is sufficiently
high during the explosion to allow  electron capture which determines $Y_e$
and, in those layers,
the initial metallicity has no influence on the final burning product
(H\"oflich et al.
1997b).

\subsection{Merging White Dwarfs}
 The second group of progenitor
models consists of two low-mass white dwarfs in a close orbit which
decays due to the emission of gravitational radiation and this, eventually,
leads to the merging of the two WDs
 (e.g. Iben \& Tutukov 1984).
After the initial merging process,
one low density WD          is surrounded by an extended envelope
 (Hachisu, Eriguchi \& Nomoto 1986,  Benz,   Thielemann \& Hills     1989).
 This scenario is mimicked by our envelope models DET2env2...6
 in which we consider the detonation in a
low mass WD          surrounded by a compact envelope between 0.2 and 0.6
$M_\odot $.

\subsection{Explosions of Sub-Chandrasekhar Mass White Dwarfs}
      Another class of models -- double detonation of a C/O-WD
triggered by detonation of helium layer in low-mass WDs
 -- was explored by Nomoto (1980),
Woosley \& Weaver (1980), and most recently by Woosley and Weaver (1994,
hereafter WW94), Livne \& Arnett (1995) and HK96.
For these models, the mass of the C/O core and
of the He-layers   given separately.

 For ease of comparison,
 we have  used parameters  close  to those suggested in WW94.
 To prevent repetition,
we refer to the latter work for a detailed discussion of this
 class of models.
 Helium detonations show a qualitatively different structure in comparison to
all models with a
 Chandrasekhar mass WD.
 The intermediate mass elements
are sandwiched by  Ni and He/Ni rich layers at the inner and outer regions,
respectively.
 Generally,  the density   smoothly decreases  with
mass because partial burning produces almost the same amount of kinetic energy
as  the total
burning, but a moderate  shell-like structure is formed just below the former
Helium layers.
 Observationally,  a distinguishing feature of this scenario
 is the presence of Helium and Ni with expansion velocities above 11,000 to
14,000 $km~ s^{-1}$.
Typically  0.07 to 0.13 $M_\odot$ of Ni are produced in the outer layers,
mainly depending on the mass of the
Helium shell.

\section{General Properties of the Explosion Models}

\begin{figure}
%\vspace{3.9cm}
\psfig{figure=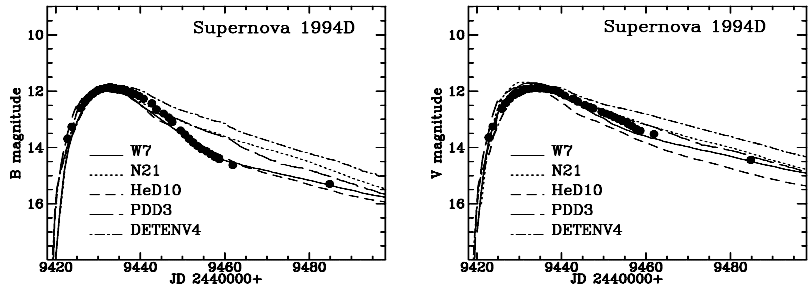,width=12.4cm,angle=270}
\caption{
Theoretical LCs of the deflagration W7, the delayed detonation
N21, the helium detonation HeD10,
 the pulsating delayed detonation  PDD3, and the envelope model DET2env4.
For comparison, the observation for SN1994D are given.
}\label{pah-4}
\end{figure}

\begin{figure}
\psfig{figure=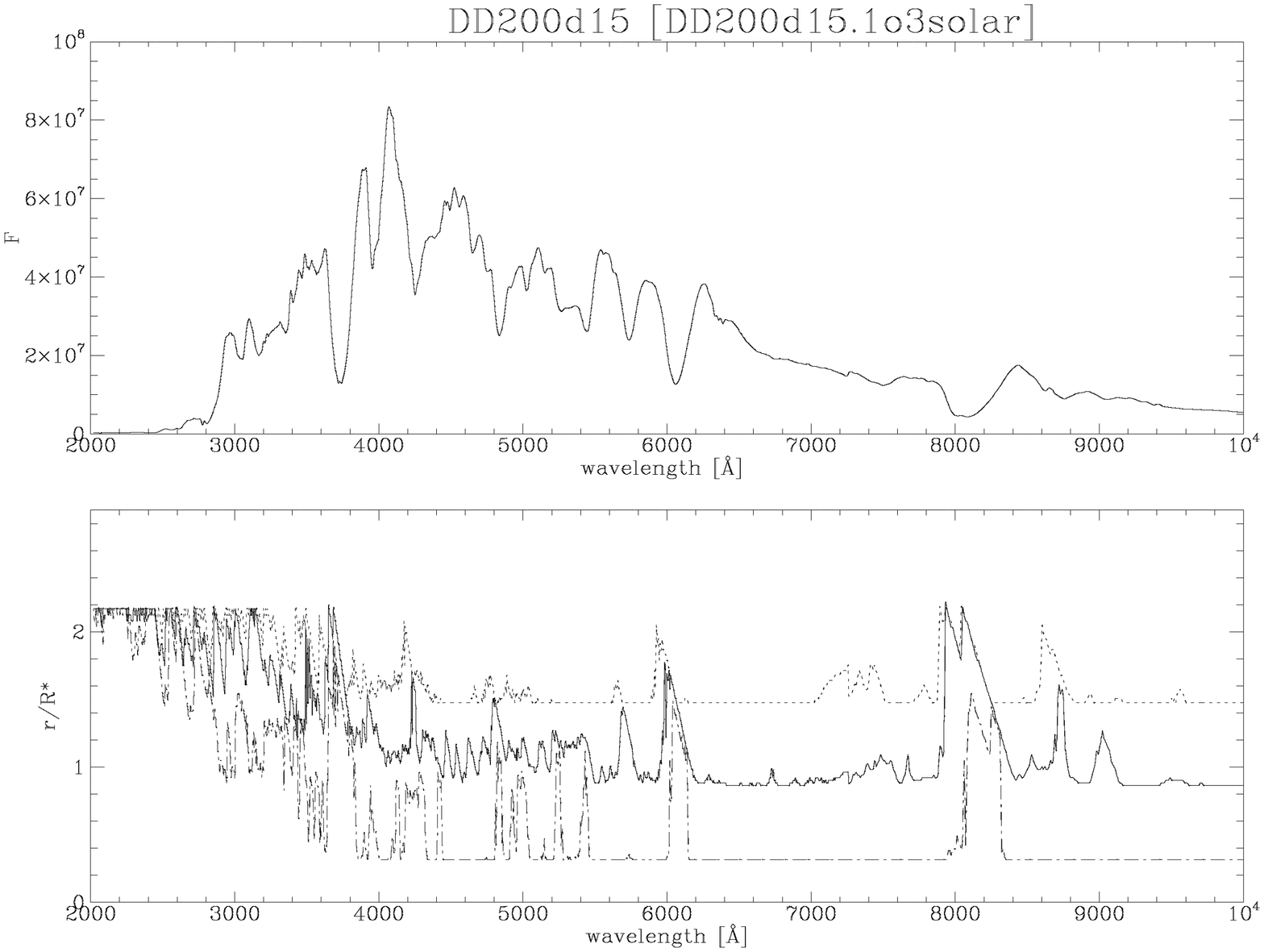,width=12.4cm,rwidth=4.4cm,clip=,angle=270}
\psfig{figure=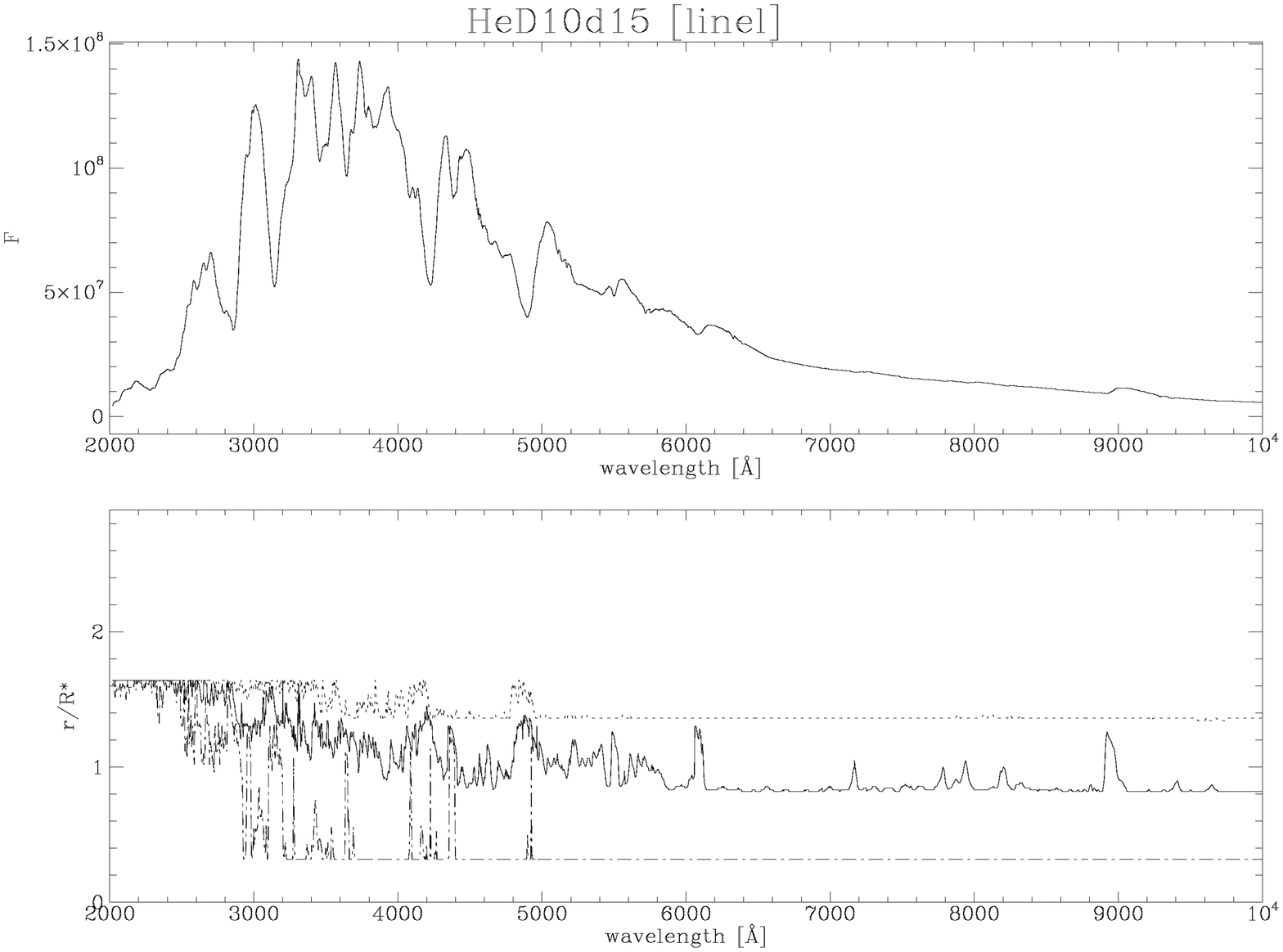,width=12.4cm,rwidth=4.4cm,clip=,angle=270}
\caption{Comparison of spectra at about maximum light of the typical delayed
detonation model
DD200 and the normal bright  helium detonation model
HeD10 (lower graph).}\label{pah-5}
\end{figure}

 The different explosion scenarios can generally be distinguished based on
differences in
the slopes of the early  monochromatic LCs (Fig.~\ref{pah-4} and spectra
(Fig.~\ref{pah-5}), e.g. by the   expansion velocities indicated by
various elements.
 Note that the differences between the slopes of the LCs are much larger
than the uncertainties imposed by the model assumptions (see sect. 2).

 For all models with a $^{56}Ni$ production  $\geq 0.4
M_\odot $, $M_V$ ranges  from -19.1 to $-19.7^m$ (HK96).
 As a general tendency, the post-maximum declines
are related to $M_V$,  but there is a significant spread in the decline rate
among models with similar
brightess.  For all models but the Helium detonations, the colors become very
red for
small  $M_{Ni}$ consistent with the obervations (Hamuy et al. 1996;
H\"oflich et al. 1996).

 The maxima of subluminous supernovae are more pronounced.
 For bright  models, the opacity  stays high and  the photosphere
recedes mainly by the geometrical dilution of matter. For models with a slower
rise time or little $^{56}Ni$, the opacity drops strongly at about maximum
light. The photosphere
receds quickly in mass, and thermal energy can be released from a larger
region.
 Because no additional energy is gained, the energy reservoir is exhausted
faster, and the
post-maximum  decline becomes steeper.

 The blue and UV spectrum of the delayed detonation model
is dominated by strongly blended lines due to Fe group elements .
 Strong features due to Si II at $\approx 6100 \AA$  and Ca II at about
$\approx  3800 \& 8100 \AA$
are consistent with observations.
 In contrast,
the spectra of Helium detonations are very blue due to the heating by
radioactive Ni both for normal
and subluminous SNe~Ia and the spectra are dominated by strong Ni lines without
showing a strong Si line.
 Already from the general properties, these HeDs can be ruled out as a general
scenario.

\section {Comparison between Observation and Models Predictions}

\begin{figure}
\psfig{figure=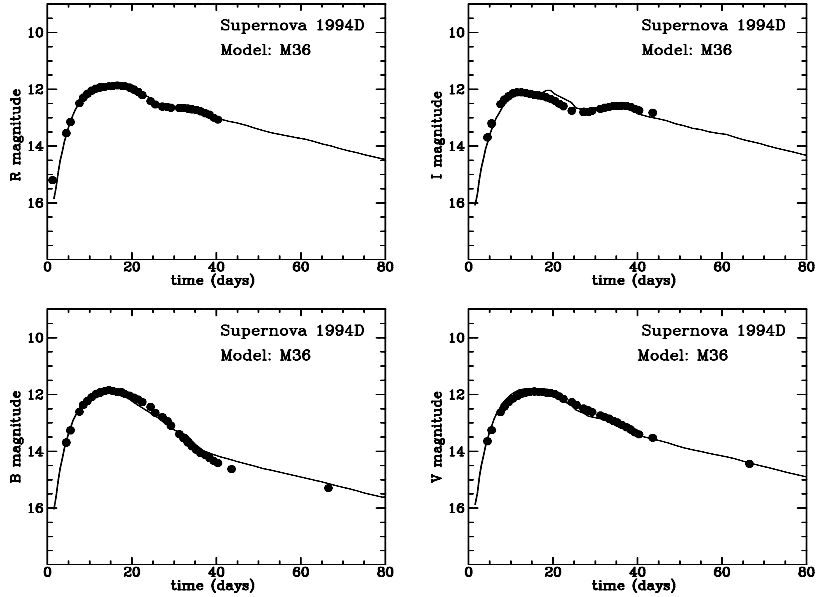,width=12.4cm,rwidth=4.4cm,clip=,angle=270}
\caption{Observed LCs of SN 1994D in comparison with the theoretical LCs of
M36}\label{pah-12}
\end{figure}

\begin{figure}
\psfig{figure=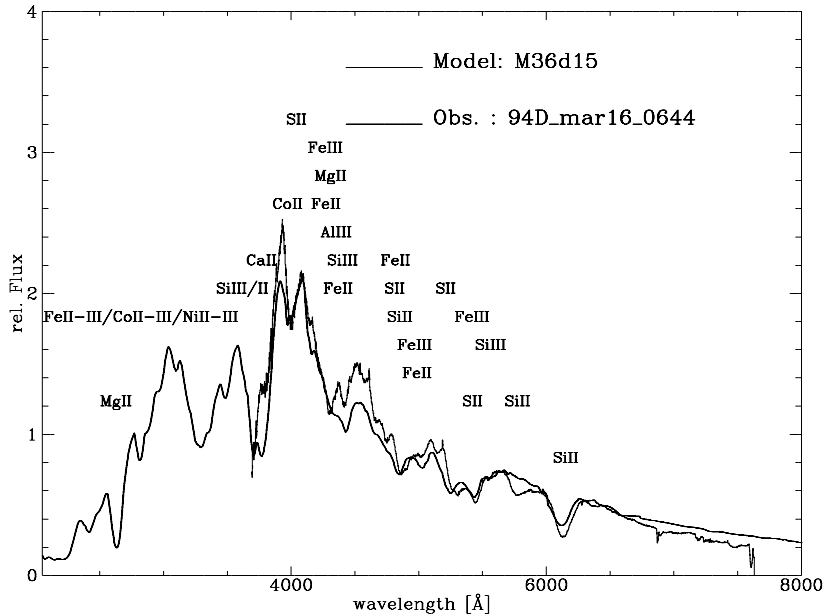,width=12.4cm,rwidth=4.4cm,clip=,angle=270}
\caption{Synthetic spectrum at day 15 for M36 compared to the observations of
SN1994D at Mar. 16th.}\label{pah-13}
\end{figure}

To get more information, detailed fitting of individual supernovae is required.
 The following 29 SNe~Ia have been analysed:
 SN~37C,  SN~70J,  SN~71G,  SN~72E,  SN~72J,  SN~73N,  SN~74G,  SN~75N,
SN~81B,  SN~83G,
 SN~84A,  SN~86G,  SN~88U,  SN~89B,  SN~90N,  SN~90T,  SN~90Y,  SN~90af,
 SN~91M,  SN~91T,  SN~91bg,  SN~92G,  SN~92K,  SN~92bc,  SN~92bo,
 SN~94D,    SN~94M,   SN~94S,   SN~94T (HK96, H\"oflich et al. 1997a,
and references therein).
For an example, see Figs.~\ref{pah-12}~\ref{pah-13}~\ref{pah-6}.
To fit the light curves,
 we use  a quantitative method for fitting  data to models based on Wiener
filtering
(Rybicki and Press, 1995).
The reconstruction technique is applied to the standard deviation from the
theoretical LC to avoid
problems with measurements  distributed   unevenly in time.
By minimizing the error,
the time of the explosion, the distance, and the reddening correction can be
determined.

According to our results, normal bright, fast SNeIa (e.g.  SN~71G,
SN~94D) with rise times up to 18~days for the
visual LC can be explained by delayed detonation with
 different densities $\rho_{tr}$ for the transition from a deflagration to
a detonation. For PDDs, the density $\rho_{tr}$ stands for the density
at which the detonation starts after the
first pulsation. Typically, $\rho _{tr} $ is about $2.5~10^7~g~cm^{-3}$.
 Central densities of the initial WDs  range from 2.1 to
   3.5 $10^9 g~cm^{-3}$. As a tendency, models at the lower end of this range
give better fits.
We note that the classical deflagration W7 (Nomoto et al. 1984)
provides similar good fits in several cases because its structure resembles
those of DD models but it has some problems with the high velocities of
Si lines in SN1994D  (H\"oflich 1995).

The ``standard" explosion models are unable, however, to reproduce very slow
rise
times ($\geq $18-20~days) even within the uncertainties,
unless the C/O ratio is rather low (see sect. 7, H\"oflich et al. 1997b), or by
models with an envelope of
typically 0.2 to $0.4M_\odot$. The envelope can be produced during a strong
pulsation
or during the  merging of two WD.
 Distinguishing features between the alternatives of low C/O ratio and an
envelope
is the presence of unburned material and the
absense of high velocity Si in the latter case, and a slow decline rate
compared to DDs
(see also sect. 7).
 Another, unique features of models
with massive envelopes are
very high photospheric expansion velocities ($v_{ph} \approx
16,000$~km/s) shortly before maximum light, which drop rapidly to an almost
constant value between 9000 and $12,000$~km/s.  This ``plateau" in
$v_{ph}$ lasts for 1 to 2~weeks depending on the envelope mass
(Khokhlov et al. 1993).  In fact, there is some evidence for the plateau in
$v_{ph}$ from the Doppler shift of lines  for SNe~Ia with  a slow
pre-maximum rise and post-maximum decline (e.g. SN~84A, SN~90N, M\"uller
\& H\"oflich 1994) and, for SN1991T, there is evidence of unburned carbon down
to velocities of 16,000 km/sec (Branch,  this volume).

Qualitatively, strongly subluminous SNe~Ia (SN~91bg, SN~92K, SN~92bc) can
be explained within the framework of pulsating
delayed detonation models with a low transition density which produce a very
low mass envelope and very little
$^{56}Ni$ but Si down to small expansion velocities (H\"oflich et al. 1995).
Due to the small heating,
these models become systematically redder and the post-maximum
decline becomes steeper with  decreasing  brightness in agreement with
observations.
 The evolution of the photospheric expansion velocity $v_{ph}$ and,
 in particular, its steady decline, is  consistent with observations.
Note that some of the `classical' delayed detonation models (M39) also produce
strongly subluminous LCs, but these do not fit  any of the measurements.

\subsection {IR-Spectra}

 Despite very interesting results from the IR,
spectroscopic analysis of early time infrared spectra of supernovae in
the literature is limited and focused on individual features. E.g.
  Meikle et al.  (1996) concentrated on
determining the origin of a feature at 1.05-$\mu$m in the spectra of
SN\,1994D. They concluded that the feature might be due to either HeI 1.083
$\mu$m
or MgII 1.0926  $\mu $m, but found difficulties with both identifications.
  Graham (1986) attributed the 1.2-$\mu$m spectroscopic feature to absorption
by
a multiplet of SiI.   Spyromilio et al. (1994)   argued that no
multiplet or combination of lines is responsible for the lack of flux
in this spectral region.   Instead, it is the absence of a line
blanketing opacity and the high transparency of the supernova in
continuum opacity (H\"oflich et al. 1993) which causes the
lack of flux in this region. However, the theoretical spectrum of Spyromilio et
al.
failed to fit the observations.

\begin{figure}
\psfig{figure=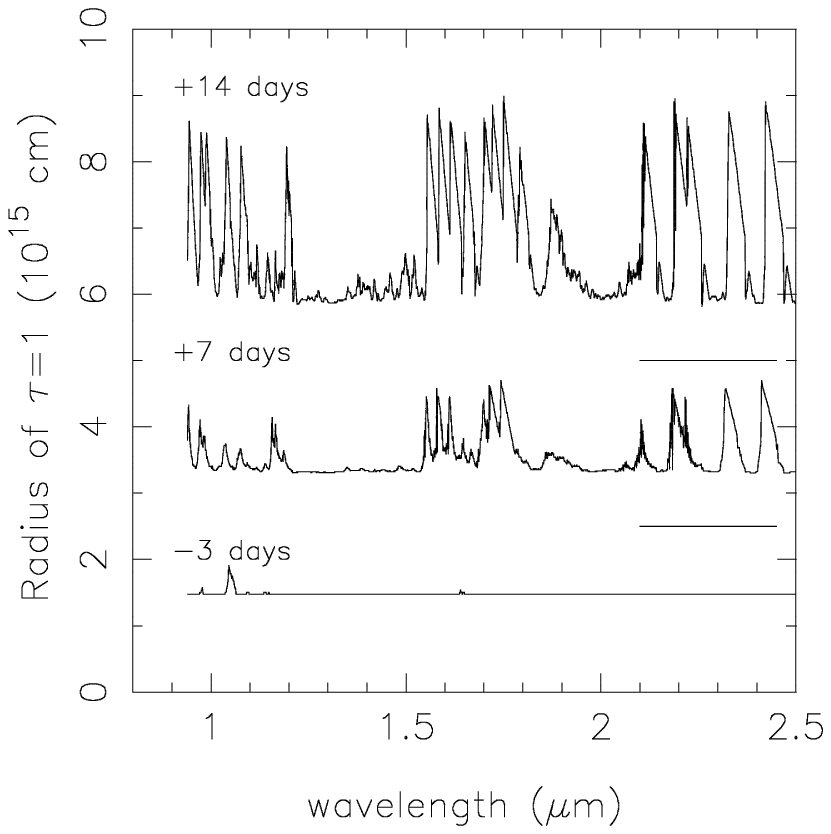,width=12.4cm,rwidth=4.4cm,clip=,angle=0}
\vskip -5.9cm
\psfig{figure=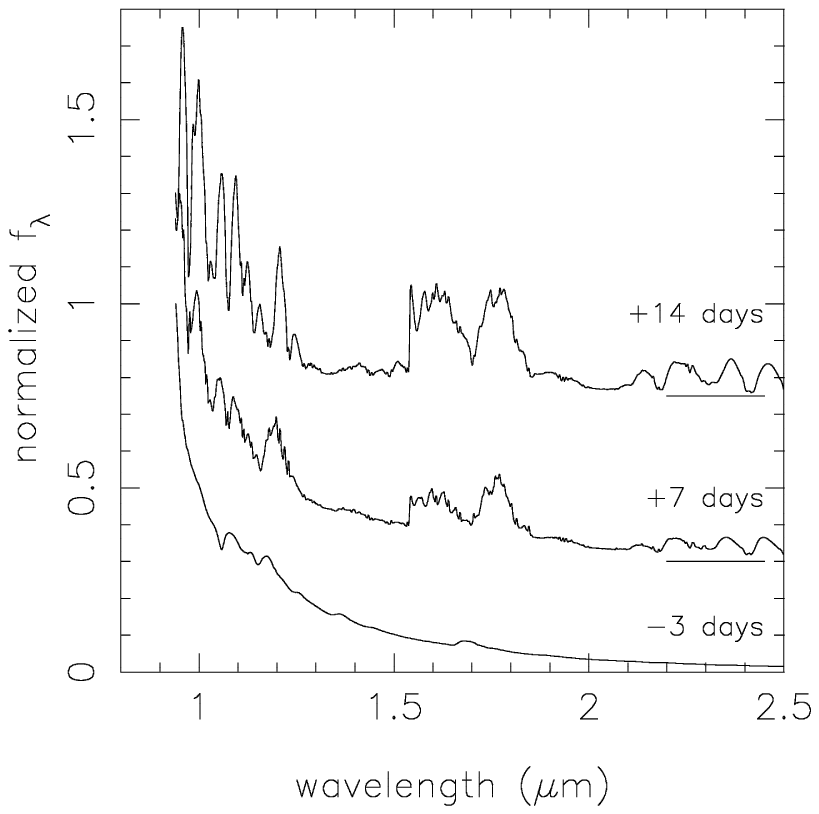,width=6.2cm,rwidth=4.4cm,clip=,angle=0}
 \caption{Evolution of the DD200 model spectrum
and the radius of optical depth unity
 (times with respect
to model V maximum).}\label{pah-6}
\end{figure}

 Here, we want to discuss the formation of the IR-spectra more general. We want
 to show the unique potential of IR-spectroscopy of
SNe~Ia, and that the current generation of models allow for  quantitative
analyses.
In Fig.~\ref{pah-6},  we give  the evolution of the spectrum  and of the radius
at
which of the optical depth is unity for DD200
are given with respect to model V maximum.  They are -3, +7 and +14 days
corresponding to about 15, 25, and 32 days after the explosion.  The spectra
have been normalized to the maximum flux in the displayed range and then a
constant has been added for clarity of display.  The levels of zero
flux  and radii are shown as straight lines in
the red part of the spectra.

Three days before V maximum, the spectrum in Fig. ~\ref{pah-6} is relatively
featureless with the exception of Mg\,II at 1.05 $\mu$m. In DDs, it is produced
as a natural consequence of burning and it is not due to He. Because Mg is
produced
mainly in the region of explosive carbon burning, the Doppler shift provides a
unique tool to determine the transition zone from explosive carbon to oxygen
burning.
Early on, the IR-opacities are dominated by electron scattering.  Later on, at
wavelengths shorter than $\sim$1.2-$\mu$m, the spectra are dominated by
blends of Mg, Ca and iron group elements.
  With increasing time, the atmosphere
becomes cooler. Once the electron scattering opacity decreases,
 a Type Ia supernova no longer has a well defined
photosphere and the spectrum forming region
varies with wavelength depending on the transition present at that
frequency interval (Fig.~\ref{pah-6}).  The features at about 1.5---1.7 and
2.2---2.6-$\mu$m are primarily due to iron group elements of the second
ionization stage.
They appear in
``emission" because they are formed at a larger radius and the source
function is not scattering dominated, but governed by the
redistribution of energy from higher frequencies (H\"oflich 1995).

Flux minima occur where the spectrum is formed at a small radius
(Fig.~\ref{pah-6}).  The broad holes in the opacity distribution may be used as
a
tool for analyzing the composition of the ejecta.
The broad holes between 1.2--1.5 and 1.9--2.1-$\mu$m correspond to a
dispersion in velocity of 70,000 and 30,000\,km\,s$^{-1}$, far
exceeding the dispersion of the expansion velocity of radioactive
material in any supernovae.  Consequently, these gaps cannot be
eliminated by velocity smearing and should in general be visible when
the continuum opacity is low enough to expose them.  The existence of
the gaps provides the opportunity to probe very different layers of the
ejecta at a given time.  This makes the IR an important complement to
the optical and the UV.

\begin{figure}
 \centering
\psfig{figure=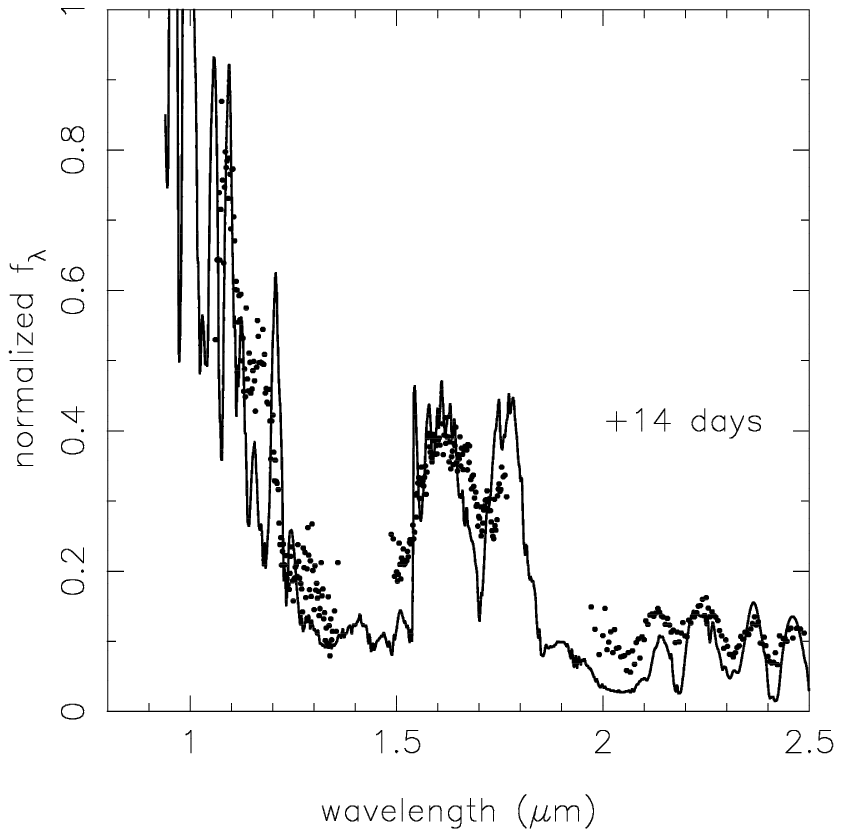,width=12.4cm,rwidth=4.4cm,clip=,angle=0}
\vskip -5.5cm
\psfig{figure=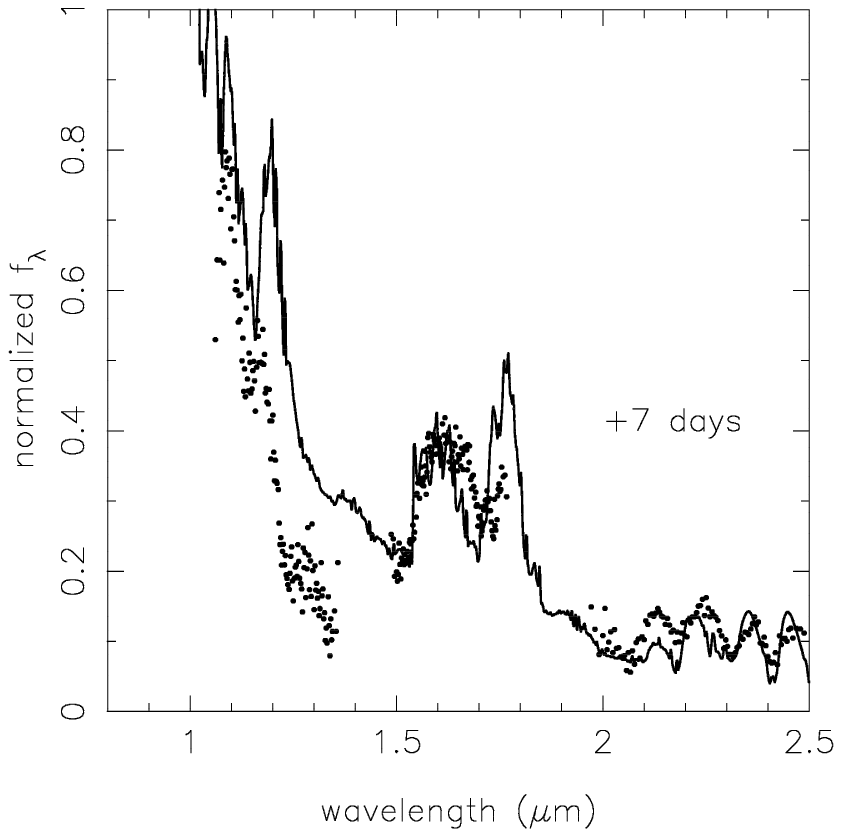,width=6.8cm,rwidth=4.4cm,clip=,angle=0}
 \caption{Synthetic spectra at +7 and +14d, and SN\,1986G at +10d
with respect to model and observed V maximum, respectively.}\label{pah-7}
\end{figure}

In Fig.~\ref{pah-7}, we compare the data from SN\,1986G with the models.
Note that SN 1986G is regarded to be a slightly subluminous, rapidly
declining event compared to ``normal" Type Ia supernovae.
We neither tuned the models to provide a best fit nor reproduced the
exact times of the observations.  Instead, the goal of the comparison is
to demonstrate that current  generation of atmosphere codes is well suited to
allow quantitative
and that the IR is a valuable tool to analyze SNe~Ia. For details, see Wheeler
et al. (1997).

\section{Distance Determinations \& $H_o$}

\begin{figure}
\psfig{figure=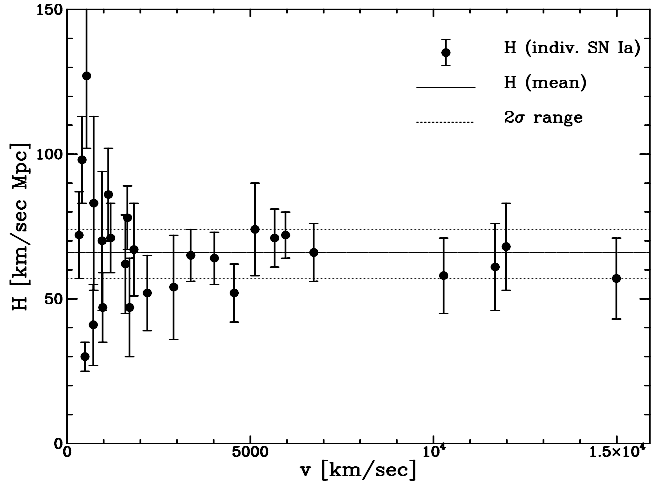,width=10.6cm,rwidth=5.0cm,angle=270}
\caption{
Hubble values H are shown based on individual distances based on fitting of the
LCs and spectra
(sect. 5 \& HK96).
}\label{pah-8}
\end{figure}

Based on our LCs, we have also determined the individual
distances of the parent galaxies of the analyzed SNe~Ia (sect. 5). Our method
does not rely on secondary distance indicators and allows for a consistent
treatment of interstellar reddening and the interstellar redshift.
 We find $H_o $ to be $67 \pm 9 km/sec Mpc$ within a 95 \% confidence level
(Fig.~\ref{pah-8}).
This value agrees well with our previous analysis based on a subset of
observations
and models ($66 \pm 10 km~Mpc^{-1} sec^{-1}$, M\"uller \& H\"oflich 1994).
 By  using the  brightness at  maximum light alone (but including the
individual
reddening correction), our $H_o$ decreases to $63  km~Mpc^{-1} sec^{-1}$.

 Other determinations of $H_o$  are based on independent,
purely statistical methods and primary distance indicators.
 It may be encouraging that the result
of different SN~Ia based methods agree
 if SN~Ia are not treated as as  standard candles. Tammann \& Sandage (1995)
found  $H_o $ of $54 \pm 4 $ and $59 \pm 4 $ using the B and V LC,
respectively.
    Hamuy et al. (1995) get
$65 \pm 5$,  Riess et al. (1995) give $67 \pm 5$,  and
Fisher et al. (1995), Nugent et al. (1995b) and Branch et al, (1995)
 get a values of $60 \pm 10$, $60 \pm 12$ and
$58 \pm 7$, respectively.
 Taking the excellent agreement between all methods, one may
conclude that the question of H$_o$ has been settled at least within an 10 \%
error range.
{}From our models, both the empirical relations between
 $M_V/dM(15)$ like-relations and the ansatz to deselect subluminous SNeIa seems
to be justified, but we expect an individual  dispersion  of $\approx 20 \%$
(H\"oflich et al. 1996).
 Therefore more detailed analysis are needed to reduce the uncertainties
beyond the current point.

\section{Evolutionary Effects and Consequences for the
Determination of $\Omega_M$ and $\Lambda$}

 Time evolution is expected to  produce the following main effects: (a) a
lower
metallicity will decrease the time scale for  stellar evolution of individual
stars
by about 20 \% from Pop I to Pop II stars (Schaller et al. 1992) and,
consequently,
the  progenitor population which contributes to the SNe~Ia rate at any given
time.
 The stellar radius also shrinks. This will influence the statistics
of systems with mass overflow;
 (b) Evolutionary effects of the stellar population will change the
mass function present at the time corresponding to a given redshift;
 (c) The initial metallicity Z will effect the nuclear burning during the
explosion or,
more precisely, the electron to nucleon fraction (sect. 3);
 (d) Systems with a shorter life time may dominate early on and, consequently,
the typical C/O ratio
 of the central region of the WD is reduced;
 (e) In principle, a change of the metallicity may alter the R-M relation of a
WD but
 this issue is not a major concern (H\"oflich et al. 1997b);
 (f) The properties of the interstellar medium may change.

Before discussing the {\sl possible (!)} implications for cosmology in more
detail,
we  address the influence of changes in the initial composition on individual
light curves and spectra, i.e. (c) and (d).
 The influence of the initial composition on light curves and spectra
 has been studied for the example of a set of delayed detonation models with
DD21c being
the reference model (Table~\ref{pahtbl-1}). Because the time evolution of the
composition is not well known and
because we have not considered the entire variety of
possible models, the values given below {\sl do not (!)} provide a basis  for
quantitative corrections
of existing observations.
 The goal is to get a first order estimate of the size of the systematic
effects to be expected,
to demonstrate how evolutionary effects in a real data sample can be recognized
and how
one may be able to compensate.

\begin{figure}
\psfig{figure=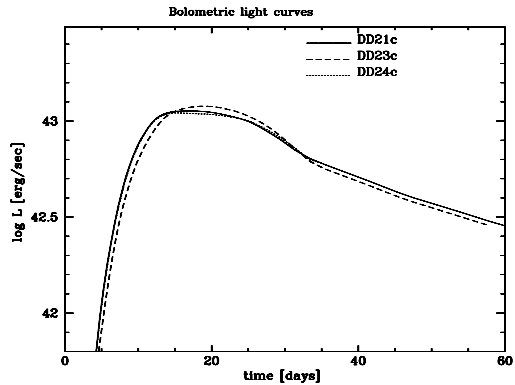,width=12.6cm,rwidth=6.5cm,clip=,angle=270}
\hsize=6.cm
\vskip -5.cm
\caption{ Comparison of bolometric LCs of the delayed detonation
models DD21c, DD23c and DD24c with otherwise identical parameters but with
different
C/O ratios and metallicity relative to solar (C/O; $R_Z$) of (1;1), (2/3;1) and
(1;0.3),
respectively.}\label{pah-9}
\end{figure}

\subsection {Light Curves}

 Reducing the ratio  C/O from 1/1 to 2/3 in the WD reduces the $^{56}$Ni
production
and the kinetic energy in DD23c compared to DD21c.
 The smaller expansion due to the smaller $E_{kin}$ causes a slower rise to
maximum light by about 3 days. The
smaller $^{56}Ni$ results in  a steeper decline after maximum light because the
photosphere recedes faster
and the luminosity after about day 35 is smaller by about 10 \%
(Fig.~\ref{pah-9}).
Both the  bolometric and monochromatic light
curves are effected by a similar amount because the energetics of the explosion
changes.
 Overall, the change of the C/O ratio
from 1/1 to 2/3 has a similar effect
on the colors, light curve shape and the distribution of elements
as a 10 \% reduction in the transition density or central density in delayed
detonations. However,
 for a given peak magnitude to tail ratio, a model produced by a reduced C/O
ratio will be brighter than a
one obtained by varying the transition density (H\"oflich 1995).
 The  rise time and the  mean expansion rate  provide a way to determine
 the C/O ratio and the transition density independently
 (compare  Khokhlov  et al., 1993, H\"oflich 1995, HK96).

 Changing the initial metallicity Z has very little influence on the bolometric
light curve
because the  $^{56}Ni$  production and energy release vary   by only 4 \% if Z
is varied between
0.1 and 10 times solar.
 In addition, diffusion time scales are mainly determined by deeper layers and,
and there, the electron
capture during burning determines $Y_e$ and not the initial composition (sect.
3).

\subsection {Spectra at maximum light}

 Spectral changes due to the C/O ratio produce a variation in the pattern of
the most abundant
elements  similar to  a change in the transition density and the central
density of the WD.
 Consequently, the variation of the spectra is similar. However, this
degeneracy can
be overcome if we combine the information from the LC with the expansion
velocity measured
by the line shifts.

\begin{figure}
\psfig{figure=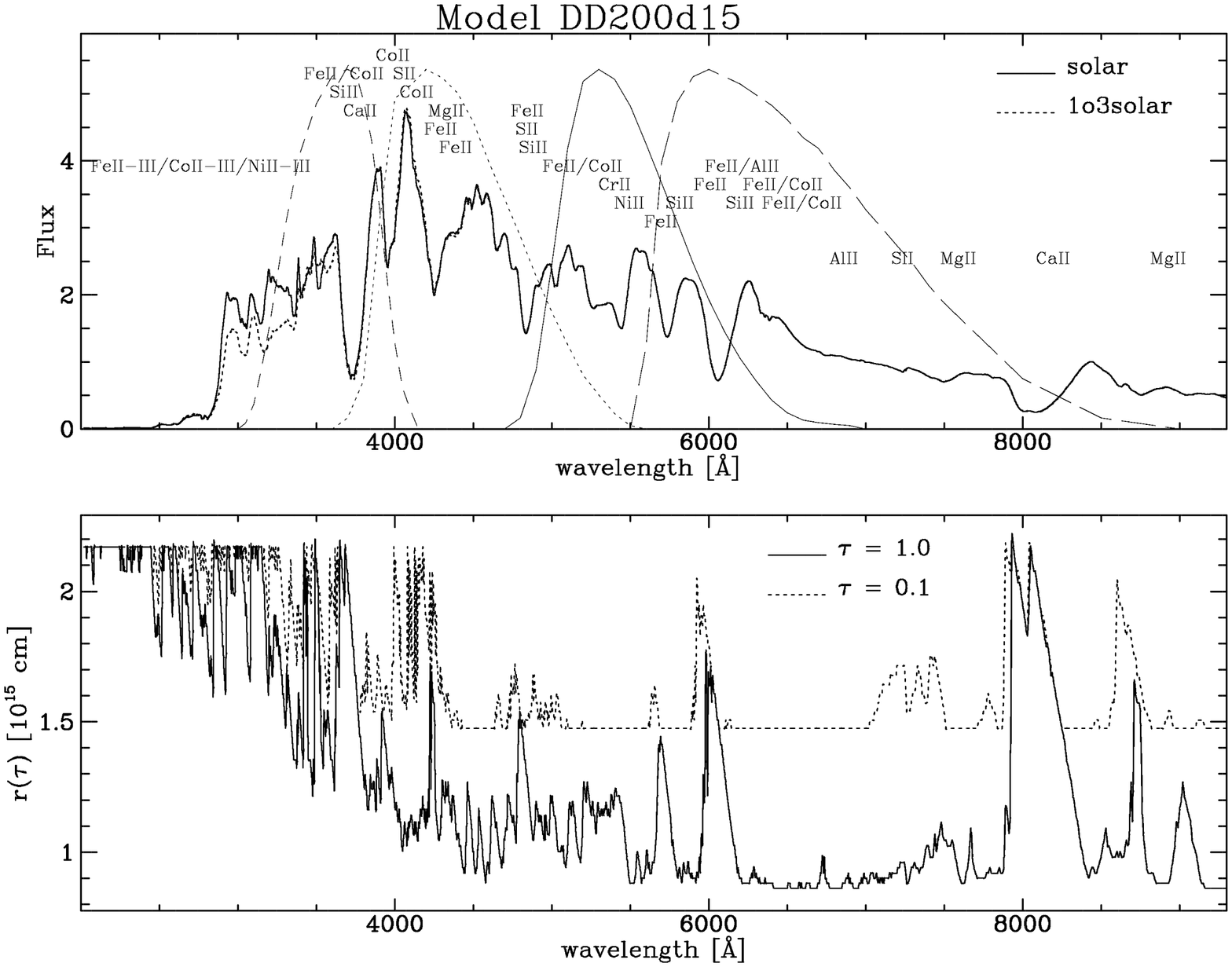,width=15.6cm,rwidth=9.5cm,clip=,angle=270}
\caption{ Comparison of synthetic NLTE spectra at maximum light for
 initial compositions of solar and 1/3 of solar, respectively.
 The standard Johnson filter functions for UBV, and R are
also shown. }\label{pah-10}
\end{figure}

 More interesting is the influence of the initial metallicity because it
influences the
iron group elements even outside the region which underwent incomplete
Si-burning during the
 explosion.
As an example, the spectra  of the delayed detonation models with solar and 1/3
solar
metallicity are given in Fig.~\ref{pah-10}. At maximum light ($\approx $ 17
days), the line forming region
($\tau \approx 0.1 ~...~1.$) extends
between 1  and 2x$10^{15}$ cm in the optical, corresponding to expansion
velocities between 8000 and
16,000 km/sec (Fig.\ref{pah-1}).  Consequently, the upper edge of the $^{56}Ni$
rich layers ($v \approx 12,000~ km/sec$) become visible at maximum light
 and all but the far UV is formed in regions where at least partial burning to
Si took place.
The spectrum is scarcely effected by  metallicity effects except for
wavelengths
below 4200 \AA~ which are dominated by the layers with $^{54}$Fe.

\subsection{Implications for Cosmology}

 Systematic errors that must be taken into account in the use of SNe~Ia to
determine
cosmological parameters include technical problems, changes
of the environment with time,  changes in the statistical
 properties of the SNe~Ia, and changes in the physical properties of SNe~Ia.

 In the first class, we put  corrections for redshift. If standard
filter systems are used, they can be well calibrated to local standards but the
k-correction is of some concern. This  problem can be overcome if redshifted
``standard" filters are used. Another
  technical problems may arise then from the fact that the
transmission functions must be identical to those resulting from the redshift
because
no direct calibration can be applied by using a comparison star.

 In the second class of problems an important example is that the properties of
dust
may change at high redshift. In the first place,  the element abundances in the
ISM can
change. In addition,
important donors  of dust such as low mass stars during the red giant phase
cannot contribute  because their evolutionary time is comparable to or longer
than
the age of the universe at z $\approx 0.5$ to $1$. Another problem related to
the correction
for extinction is that
the extinction law to be applied depends on the redshift
of the absorbing dust cloud (HK96).

 In the  third class of problems is the fact  that the contribution of
different
progenitor types may change with redshift.
 For local supernovae, it is likely that a large variety of binary
star properties (total mass of donor stars, separation, etc.) with very
different
 evolutionary life times
 account for the variety of SNe~Ia observed (Canal et al. 1997, H\"oflich et
al. 1997a).
Given this variety, we  expect a time evolution of
the statistical properties of the progenitors because, early on, progenitors
with
a short life time will dominate the sample. Among other changes, more massive,
shorter lifetime progenitors
will  have a lower C/O ratio in the center. Changes in the statistical
properties
are expected to increase with redshifts larger than 0.7 to 0.8 when the age of
the univers becomes
comparable or shorter than the suspected  progenitor life times.
 Other evolutionary effects have already been
mentioned: The ZAMS life time changes with metallicity,  Roche lobe radii
change with
metallicity and  the lower limit for accretion and steady burning of hydrogen
 on the surface of the  WD
changes by a factor of 2 for Pop I and Pop II stars, respectively (Nomoto et
al., 1982).

Finally, the physical properties of a typical SNe~Ia may change. If more
massive stars contribute
to the supernovae population, we expect a smaller mean C/O ratio. Changing the
 C/O ratio from 1/1 to  2/3 with otherwise identical parameters (see sect.
7.1),
 will result in a smaller $^{56}Ni$ production and, consequently, a lower
bolometric luminosity at late times.
 On the other hand, the slower expansion causes less adiabatic cooling during
early times and, thus,
the luminosity at maximum light is larger. This implies that the peak to tail
luminosity
ratio  changes. The consequences for the analysis by using light curve shapes
or the postmaximum decline
(e.g. $\delta m_{15}$) is evident. Quantitatively, for monochromatic light
curves in our example,
the difference in the peak to tail luminosity ratio
is $\approx ~0.2 ^m$. For techniques to determine the absolute peak luminosity
by measuring either  $\delta m_{15}$ or LC shapes (e.g. Hamuy et al. 1996, Ries
et al. 1995),
 this translates
into a systematic error of $\approx 0.3^m$ in $m_v$. This systematic change in
peak luminosity with
redshift due to changing C/O ratios could mimic the effects of cosmology.
Note that the transition density from
deflagration to detonation may depend on the energy release during the
deflagration phase and, hence
the transition density could be a function of the C/O ratio, both effects could
alter the LC.

\begin{figure}
\hsize=6.cm
\psfig{figure=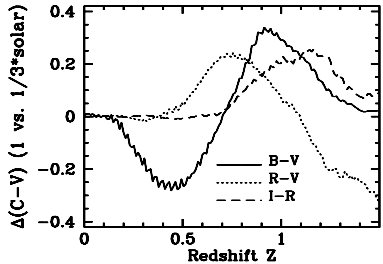,width=15.6cm,rwidth=10.5cm,clip=,angle=270}
\vskip -3.2cm
\caption{Change of  B-V, R-V, and I-R as a function of the
redshift for DD21 when changing the initial abundance from solar to
1/3*solar.}\label{pah-11}
\end{figure}

The other physical effect on the SNe~Ia themselves
 is the influence of the initial metallicity Z on the nuclear
 burning conditions during the explosion.  This
 produces a change of the isotopic composition in the outer layers. A reduction
of Z with redshift z
is expected.
 Although small, the changes in the spectrum with metallicity
 have an important effect on the colors of SN at high red shifts where they are
shifted into other
bands (Fig.~\ref{pah-11}.
 For local SNe~Ia, a change of the metallicity by a factor of 3 implies a
variation in color of
two to three hundredths of
a magnitude. However, if the short wavelenght part is shifted into the filter
bands which are used,
systematic effects of the order of several tenth of a magnitude can be
expected. When using the
color indices B-V, R-V and R-I, the systematic effects may be expected for z
$\ge $ 0.2, 0.5 and 0.7,
respectively.

\acknowledgments
 Its a pleasure to thank all my collaborators and
to thank  G.A. Tammann and his (extended) group
  for the interesting discussions during my stay at the
University of Basel.

\end{document}